\documentstyle[11pt]{article}
\def\Im{\mathop{{\cal I}\mskip-4.5mu \lower.1ex \hbox{\it m}}}
\def\Re{\mathop{{\cal R}\mskip-4mu \lower.1ex \hbox{\it e}}}
\def\simgt{\rlap{\lower 3.5 pt \hbox{$\mathchar \sim$}} \raise 1pt \hbox {$>$}}
\def\simlt{\rlap{\lower 3.5 pt \hbox{$\mathchar \sim$}} \raise 1pt \hbox {$<$}}
\def\hateq{\widehat{=}}

\def\bt{\beta_\tau}
\def\bpi{\beta_\pi}
\def\cm{\cos \theta_-^*}
\def\cp{\cos \theta_+^*}
\catcode`\@=11

\def\section{\@startsection{section}{1}{\z@}{3.5ex plus 1ex minus .2ex}
{2.3ex plus .2ex}{\large\bf}}

\def\thesection{\arabic{section}.}

\def\appendix{\setcounter{section}{0}
 \def\thesection{Appendix \Alph{section}:}
 \def\theequation{\Alph{section}.\arabic{equation}}}

\def\citenum#1{{\def\@cite##1##2{##1}\cite{#1}}}
\def\citea#1{\@cite{#1}{}}
\def\@citex[#1]#2{\if@filesw\immediate\write\@auxout{\string\citation{#2}}\fi
  \def\@citea{}\@cite{\@for\@citeb:=#2\do
    {\@citea\def\@citea{,\penalty\@m}\@ifundefined
       {b@\@citeb}{{\bf ?}\@warning
       {Citation `\@citeb' on page \thepage \space undefined}}%
\hbox{\csname b@\@citeb\endcsname}}}{#1}}

\def\citer{\@ifnextchar [{\@tempswatrue\@citexr}{\@tempswafalse\@citexr[]}}

\def\@citexr[#1]#2{\if@filesw\immediate\write\@auxout{\string\citation{#2}}\fi
  \def\@citea{}\@cite{\@for\@citeb:=#2\do
    {\@citea\def\@citea{--\penalty\@m}\@ifundefined
       {b@\@citeb}{{\bf ?}\@warning
       {Citation `\@citeb' on page \thepage \space undefined}}%
\hbox{\csname b@\@citeb\endcsname}}}{#1}}

\def\figitem{\@ifnextchar[{\@lfigitem}{\@figitem}}
\def\@lfigitem[#1]#2{\item{Fig.~\@figlabel{#1}.}\if@filesw
      { \def\protect##1{\string ##1\space}\immediate
        \write\@auxout{\string\figcite{#2}{#1}}\fi\ignorespaces}}
\def\@figitem#1{\item\if@filesw \immediate\write\@auxout
       {\string\figcite{#1}{\the\c@enumi}}\fi\ignorespaces}
\def\figcite#1#2{\global\@namedef{fb@#1}{#2}}
\let\citation\@gobble

\def\fcite{\@ifnextchar [{\@tempswatrue\@fcitex}{\@tempswafalse\@fcitex{[]}}}
\def\@fcitex#1#2{\if@filesw\immediate\write\@auxout{\string\citation{#2}}\fi
  \def\@fcitea{}\@fcite{\@for\@fciteb:=#2\do
    {\@fcitea\def\@fcitea{,}\@ifundefined
       {fb@\@fciteb}{{\bf ?}\@warning
       {Figure `\@fciteb' on page \thepage \space undefined}}%
\hbox{\csname fb@\@fciteb\endcsname}}}{#1}}
\let\figdata=\@gobble
\let\figliststyle=\@gobble
\def\figlist#1{\if@filesw\immediate\write\@auxout{\string\figdata{#1}}\fi
  \@input{\jobname.figs}}
\def\figliststyle#1{\if@filesw\immediate\write\@auxout
    {\string\figliststyle{#1}}\fi}

\def\@fcite#1#2{#1\if@tempswa , #2\fi}

\def\thefiglist#1{
 \par\clearpage\section*{Figures \@mkboth{Figures}{Figures}}
\list
   {\m@th{Fig.\ \arabic{enumi}.\ \hfill}}
   {\settowidth\labelwidth{Fig.\ \m@th {#1}.\ }%
    \leftmargin\labelwidth
    \advance\leftmargin\labelsep
    \usecounter{enumi}}
    \def\newblock{\hskip .11em plus .33em minus -.07em}
    \sloppy
    \sfcode`\.=1000\relax}

\relax

\begin{document}

 \hfill\vbox{\hbox{\bf DESY 93--174}
 	    \hbox{\bf TTP 93--41}
           \hbox{December 1993}}
\vspace{0.5in}
\begin{center}
{\large\bf PROSPECTS OF MEASURING THE PARITY OF HIGGS PARTICLES  } \\
\vspace{0.5in}
M.~Kr\"amer$^a$, J.~K\"uhn$^b$,
M.~L.~Stong$^b$, and P.~M.~Zerwas$^c$ \\
\vspace{0.5in}
$^a$Inst.~Physik, Johannes Gutenberg--Universit\"at, D-55099 Mainz FRG\\
$^b$Inst.~Theor.~Teilchenphysik, Univ. Karlsruhe, D-76128 Karlsruhe FRG\\
$^c$Deutsches Elektronen--Synchrotron DESY, D-22603 Hamburg FRG\\
\end{center}
\vspace{0.5in}

\begin{center}
ABSTRACT \\
\end{center}

We analyze the prospects of measuring the parity of Higgs particles
in the Standard Model and its supersymmetric extensions.  Higgs
decays are discussed in this context as well as production processes
including, in particular, the fusion  of Higgs particles in linearly
polarized photon--photon collisions.
\newpage

\section{Introduction}

While the Higgs particle in the Standard Model \cite{higgs} must
necessarily be a scalar particle, assigned the external quantum
numbers $J^{PC} = 0^{++}$, the Higgs spectrum in extended models
such as supersymmetric theories includes also pseudoscalar $0^{-+}$
states \cite{gh88}. This non-trivial assignment of the quantum
numbers invites the investigation of experimental opportunities to
measure the parity of the Higgs states. This problem was approached
in a general form quite early in Ref.~\cite{dell89} for Higgs
decays into fermion and gauge boson pairs. It has recently been
revived in a more specific form, and including production processes,
for neutral particles in Refs.~\cite{bar94,hs94} and for charged
particles in Ref.~\cite{hm}.

\vspace{3mm}

Spin correlations of fermions from Higgs decay (e.g., $\tau^+ \tau^-$
or $ t \bar t\,$) are also sensitive to the parity of the Higgs \cite{nel}.
This sensitivity is reflected in angular correlations of secondary
decay products, in particular in the acollinearity distribution of
$\pi^+ \pi^-$ from $\tau$ decays or $l^+ l^-$ from top decays.

\vspace{3mm}

Another interesting method is provided by the fusion of
neutral Higgs particles \citer{gh93,bor93}
in linearly polarized photon--photon
collisions \cite{gg92}. Polarized high energy photon beams can be
generated by Compton back-scattering of linearly polarized laser light
\cite{ginz84}. The production of scalar particles requires parallel
polarization of the two photons involved, while pseudoscalar particles
require perpendicular polarization vectors \cite{yang49}.

\vspace{3mm}

Extracting the Higgs signal in $\gamma \gamma$ collisions is not
an easy task since large numbers of $b\bar{b}/c\bar{c}$ and
$W^+W^-/ZZ$ background events must be rejected
\cite{gh93,halzen93,jikia93}. However,
the resolved $\gamma$ mechanisms do not pose background problems in
the kinematical configurations relevant to asymmetry measurements.
Choosing the laser energy only slightly higher than the
Higgs threshold in the $\gamma\gamma$ collisions requires, the
resolved $\gamma$ processes $\gamma g \to b\bar{b}$ and
$gg \to b\bar{b}$ are strongly
suppressed due to the steeply falling gluon spectrum. Background
$\gamma\gamma, \gamma g, gg \to c\bar{c}$ charmed particle events
must be suppressed by means of excellent $\mu$--vertex detectors.

\vspace{1cm}

\section{Neutral Higgs Decays to Fermion Pairs}

The most frequent fermion decay mode of neutral Higgs particles
is in general the $b\bar{b}$ channel, Fig.\ \fcite{f_higgsdec}.
This applies in the Standard Model ($\cal{SM}$) \cite{zerwas1,kniehl}
as well as its minimal supersymmetric extension ($\cal{MSSM}$)
\cite{zerwas2}. However, due to the depolarization effects in the
fragmentation process, it is very difficult to extract information
on the $b$ polarization state \cite{mele}.  Much cleaner channels,
though with branching ratios suppressed by an order of magnitude,
are the $\tau$ and $t$ modes \footnote{The generic notation $H$ will
be used for the $0^{++}$ Higgs particles, while $A$ will denote
the pseudoscalar $0^{-+}$ Higgs particles.}
\begin{eqnarray} \nonumber
H, A & \rightarrow & \tau^+ \tau^- \label{eq:tau_mode} \\
H, A & \rightarrow & t \bar{t} \label{eq:top_mode}
\end{eqnarray}
The $\tau$ channel is useful in the $\cal{SM}$ for Higgs masses
less than $\sim 130$ GeV, in supersymmetric theories  generally
over a much larger mass range \cite{zerwas2}.  Top quark decays are
of interest in a wide range above the top threshold \cite{ha91}.
For large top masses, the top quarks decay, $t \rightarrow b W$,
before fragmentation destroys the $t$-spin information \cite{bigi81}.
The top quarks can thus be treated as undressed particles, like leptons.

\vspace{3mm}

Denoting the spin vectors of the fermion $f$ and the antifermion
$\bar{f}$ in their respective rest frames by $s$ and $\bar{s}$,
respectively, [the $\hat{z}$-axis oriented in the $f$ flight
direction], the spin dependence of the decay
probability is given by \cite{bar94}
\begin{eqnarray}
\Gamma(H,A \rightarrow f \bar{f}) & \sim & 1 - s_z \bar{s}_z
		\pm s_\perp \bar{s}_\perp  \label{eq:ff_spins}
\end{eqnarray}
This spin dependence translates directly into correlations among
the fermion decay products.  A few representative examples will
be discussed in more detail.

\vspace{1cm}

\subsection{ $H,A \rightarrow \tau^+ \tau^-
            \rightarrow \pi^+ \overline{\nu} \pi^- \nu$}

Even though the decay mode $\tau^\pm \rightarrow \pi^\pm
\nu_\tau (\overline{\nu}_\tau)$ is rare, it may serve as
a simple example that illustrates the basic principles.
Defining the polar angles between the $\pi^\pm$ and the
$\tau^-$ direction in the $\tau^\pm$ rest frames by
$\theta^*_\pm$ and the relative azimuthal angle $\phi^*$
between the decay planes (Fig.~\fcite{f_tau_angdist}),
the angular correlation may be written \cite{tau_semilept}
\begin{eqnarray}
{1 \over \Gamma} { d \Gamma (H, A \rightarrow \pi^+ \overline{\nu} \pi^- \nu)
\over d \cp d \cm d \phi^*} & = & {1 \over 8 \pi} \left[ 1 + \cm
\cp \mp \sin \theta^*_+ \sin \theta^*_- \cos \phi^* \right]
\label{eq:pi_dist}
\end{eqnarray}
Anticipating that the Higgs masses will be known very accurately
before measurements of the parity will be carried out, the Higgs
rest frame can in principle be reconstructed on an event-by-event
basis 
in the Higgs--strahlung process $e^+ e^- \rightarrow Z H$ and for
associated production $e^+ e^- \rightarrow Z A$.  In practice,
measurement errors and $\gamma$ radiation will smear out the
reconstructed momenta of the final-state particles.  Further
complications occur in the reconstruction of the $\tau$ direction
of flight, as the neutrinos are undetected. The angles $\theta^*_\pm$
are related to the pion energies in the Higgs rest frame and they
can therefore be measured, but a two-fold ambiguity remains in the
$\tau$ direction and hence in the angle $\phi^*$. This ambiguity
can in principle be resolved using the impact parameter between
the pion momenta \cite{tau_recons}.  In our case, however, this
procedure is unnecessary, as only $\cos \phi^*$ is required which
may simply be extracted from the dot product of the two pion
momenta in the Higgs rest frame:
\begin{eqnarray}
16 \vec \pi_+ \cdot \vec \pi_- & = &
	m_H^2 \left( \bpi \cm + \bt \right)
	\left( \bpi \cp - \bt \right)
	+ 4 m_\tau^2 \sin \theta^*_+ \sin \theta^*_- \cos \phi^*
\label{eq:pi_dotprod}
\end{eqnarray}
$\bt = \sqrt{ 1 - 4m_\tau^2/m_H^2 \,}$ and
$\bpi = (m_\tau^2 - m_\pi^2)/(m_\tau^2 + m_\pi^2)$ are the
Higgs--to--$\tau$ and $\tau$--to--pion rest frame boosts,
respectively.
This allows us to reconstruct the distribution (\ref{eq:pi_dist})
directly.

\vspace{3mm}

A simple asymmetry in the azimuthal angle
that projects out nicely the parity of the particle,
can be derived \cite{dell89,bar94} by integrating out the polar angles
\begin{eqnarray}
{1 \over \Gamma} {d \Gamma(H,A) \over d \phi^*} & = &
{1 \over 2 \pi} \left[ 1 \mp {\pi^2 \over 16} \cos \phi^* \right]
\label{eq:pi_asymm_phi}
\end{eqnarray}

\vspace{3mm}

A useful observable sensitive to the parity of the decaying
Higgs particle is the angle $\delta$ between the two charged
pions in the
Higgs rest frame \cite{tau_semilept}.  The distribution of
eqn.(\ref{eq:pi_dist}) is rewritten in terms of the angle $\delta$
through the substitution of the pion momenta in
eqn.(\ref{eq:pi_dotprod}):
\begin{eqnarray}
16 \vec \pi_+ \cdot \vec \pi_- &=& m_h^2
\left[ (1 + \bt \bpi \cm)^2  - 16 m_\pi^2/m_h^2 \right ]^{1/2}
\nonumber \\ && \qquad \qquad
\times \left[ (1 - \bt \bpi \cp)^2  - 16 m_\pi^2/m_h^2 \right ]^{1/2}
\cos \delta
\label{eq:pidot_energies}
\end{eqnarray}
The azimuthal angle $\phi^*$ can therefore be written
as a function of the angles $\theta_\pm^*$ and $\delta$:
\begin{eqnarray} \displaystyle
{1 \over \Gamma} {d \Gamma(H,A) \over d \cm d \cp d \cos \delta} & = &
{ J \over 8 \pi } \left[ 1 + \cm \cp \mp \sin \theta^*_+ \sin \theta^*_-
\cos \phi^*(\theta^*_+,\theta^*_-,\delta) \right]
\label{eq:pi_asy_delta}\nonumber \\
J &=& { 8 \vec \pi_+ \cdot \vec \pi_- \over m_\tau^2 \cos \delta
	\sin \theta^*_+ \sin \theta^*_-
	\sin \phi^*(\theta^*_+,\theta^*_-,\delta)}
\label{eq:pi_delta_jacobean}
\end{eqnarray}

\vspace{3mm}

Integration over the polar angles $\theta^*_\pm$ gives
a rather complex function of $\delta$.  Although the
resulting distributions are very similar for most values
of $\delta$, they behave differently in the limit
$\delta \rightarrow \pi$. The scalar distribution approaches
its maximum at $\delta = \pi$,
\begin{eqnarray}
{1 \over \Gamma_H} {d \Gamma (H) \over d \cos \delta} & \rightarrow &
{2 \over 15}{\left( 5 + \bt^2 \right)
	\over \left( 1 - \bt^2   \right)},
\label{eq:pi_dist_sc_lim}
\end{eqnarray}
while the pseudoscalar distribution peaks at a small but
nonzero value of $\pi -  \delta$, corresponding to a
nonvanishing acollinearity angle between the pions.  In
the limit of vanishing pion mass, the distribution approaches
zero as the pions are emitted back-to-back:
\begin{eqnarray}
{1 \over \Gamma_A} {d \Gamma (A) \over d \cos \delta} & \rightarrow &
\left( 1 + \cos \delta \right)
{1 \over 20}{\left( 5 + 10 \bt^2 + \bt^4 \right)
	\over \left( 1 - \bt^2   \right)^2}
\label{eq:pi_dist_ps_lim}
\end{eqnarray}
Since the pion mass is very much smaller than the $\tau$
mass, the distributions for non--zero pion masses have much
the same behavior in the limit of back-to-back pions,
Fig.~\fcite{f_pidist}.

\vspace{3mm}

The discussion of $\tau$ decays to multipion final states
follows very much the same line.  One may either treat the
hadron system as a single particle with definite spin and
variable mass $\sqrt{Q^2}$, or one may extract additional
information from the individual pion momenta. The first
strategy leads to a trivial generalization of the single pion
case, the second will be detailed below.

\vspace{3mm}

Let us for definiteness consider the case of both $\tau$'s
decaying to $\rho ( \hateq 2 \pi )$.  The correlation term
in eqn.(\ref{eq:pi_dist}) is then reduced by the factor
$(m_\tau^2 - 2 Q^2)^2/(m_\tau^2 + 2 Q^2)^2$. The kinematic
relations must also be corrected by the mass of the hadron
system $Q^2$ which can no longer be neglected relative to
$m_\tau^2$.  Predictions for the distribution of the acollinearity
are shown in Fig.\ \fcite{f_rhodist} for fixed
$Q^2 = m_\rho^2 = (0.770~\mbox{GeV})^2$.

\vspace{3mm}

The dilution factor, which is even more severe in the
three-pion channel with $Q^2 \approx m_\tau^2/2$, can be
circumvented by the analysis of the individual pion
distributions.  The direction of the pion momentum (defined
in the $\tau$ rest frame) appears in eqn.(\ref{eq:pi_dist})
and replaces the spin vector $\vec{s}$ in eqn.(\ref{eq:ff_spins}).
In the general case, $s (\bar{s}$) must be replaced by the vector
$\pm R^\mp/(m_\tau \omega_\mp)$, where $R$ and $\omega$ are
defined by \cite{tau_semilept}
\begin{eqnarray}
\Pi_\mu & = & 4 \Re J_\mu q \cdot J^* - 2q_\mu J \cdot J^*
\label{eq:pimu_def}\nonumber \\
\Pi_{5 \mu} & = & 2 \epsilon_{\mu \nu \rho \sigma}
	\Im J^\rho J^{* \nu} q^\sigma
\label{eq:pi5_def}\nonumber \\
\omega & = & p_\mu ( \Pi^\mu - \gamma_{\mbox{{\scriptsize AV}}} \Pi^\mu_5 )
\label{eq:omega_def}\nonumber \\
R_\mu & = & (m_\tau^2 g_{\mu \nu} - p_\mu p_\nu)
	( \gamma_{\mbox{{\scriptsize AV}}} \Pi^\nu - \Pi^\nu_5 )
\label{eq:r_def}
\end{eqnarray}
$q$ is the momentum of the neutrino, $p_\pm$ is the momentum
of the $\tau^\pm$ and $J_\pm$ is the hadronic current, and
$\gamma_{\mbox{{\scriptsize AV}}} = 2 g_{\mbox{{\scriptsize A}}}
g_{\mbox{{\scriptsize V}}} / (g_{\mbox{{\scriptsize A}}}^2
+ g_{\mbox{{\scriptsize V}}}^2)$.
If the $\tau$ rest frame is reconstructed, for example with the
help of microvertex detectors, then $\vec R/(m_\tau \omega)$ can
be evaluated in this frame. Since $ R_\mu R^\mu /(m_\tau \omega)^2 = - 1$,
the sensitivity is completely retained in this case.

\vspace{1cm}

\subsection{ $H,A \rightarrow t \bar{t} \rightarrow (b W^+) +
            (\bar{b} W^-)$}

This partonic final state can be treated in direct analogy to
the pionic two-particle $\tau$ decay.  Including the non-zero
mass effects, the angular distribution of the $W^\pm$ bosons
is given by
\begin{eqnarray} \nonumber
&&{1 \over \Gamma} { d \Gamma (H, A \rightarrow W^+ W^- b {\bar b}) \over
d \cp d \cm d \phi^*}  \nonumber \\
&&\qquad\qquad\qquad
 = {1 \over 8 \pi} \left [ 1 + {\left( m_t^2 - 2 m_W^2 \right)^2
			\over \left( m_t^2 + 2 m_W^2 \right)^2}
	\left[ \cp \cm \mp
	\sin \theta^*_+ \sin \theta^*_- \cos \phi^* \right]\right ]
\label{eq:w_dist}
\end{eqnarray}
$\theta^*_\pm$ denote the $W^\pm$ polar angles, $\phi^*$ the
azimuthal angle between the decay planes, Fig.\ \fcite{f_w_angdist}.
The kinematical reconstruction will be much simpler in this case
than for $\tau$ decays -- despite the jetty character of the
final state -- as the mass ratio of daughter-to-parent particles
is much larger and the decay polar angles are not driven to $0,\pi$
by Lorentz boosts.  The angle $\phi^*$ is again determined only up to a
two-fold ambiguity, but as before this ambiguity does not affect the
distributions.

\vspace{3mm}

A particularly interesting process is provided by subsequent
decays of the $W^\pm$ bosons to leptons.  In this case too, the
top quark direction can be reconstructed completely. The
distribution obtained after integration over the $b$-quark
directions is exactly the same as in eqn.(\ref{eq:pi_dist})
with the angles $\theta^*_\pm$ denoting now the polar angles
between the leptons and the top quarks in the quark rest frames.
The form is similar to the distribution of the pions from
the $\tau$ decay mode, and thus without the suppression factor
$((m_t^2 - 2 m_W^2)/(m_t^2 + 2 m_W^2))^2$.
Furthermore, the difference between scalar and pseudoscalar
distributions is visible over a much larger angular range,
as the Higgs--to--top boosts are generally small, see
Fig.\ \fcite{f_leptdist}.

\vspace{1cm}

\section{Higgs Decays to  $W$, $Z$  Pairs}

Experimental opportunities to measure the parity of the Higgs
particle in the decay processes
\begin{eqnarray}
H, A \rightarrow W^+ W^- \; \mbox{and} \; ZZ \label{eq:boson_pairs_mode}
\end{eqnarray}
have been investigated earlier \cite{dell89,bar94}.
Nevertheless, for the sake of completeness, the main points
are summarized here.

\vspace{3mm}

Even though the coupling of gauge bosons to the pseudoscalar state
$A$ is not forbidden by any fundamental principle [as the decay
$\pi^0 \rightarrow \gamma \gamma$ tells us], the coupling
vanishes at the Born level and proceeds only through
triangular loop effects in two-doublet Higgs models \cite{mendez}.
However, independently of the low rate, this decay mode is useful
theoretically by allowing us to check the uniqueness of the predictions
for $0^{++}$ decay distributions to the $W$, $Z$ bosons.

\vspace{3mm}

Vector particles couple to scalar Higgs particles in S waves,
$\sim \epsilon_1 \cdot \epsilon_2$, and they couple
to pseudoscalar Higgs particles in P waves,
$\sim (\vec \epsilon_1 \times \vec \epsilon_2) \cdot \vec k_V$.
While the transverse polarization vectors are independent of the
energy, the longitudinal polarization vectors grow with the energy
of the particle.
As a result, the vector particles are polarized longitudinally
for asymptotic energies in the $0^{++}$ case, but they are polarized
transversely, independently of the energy, in the $0^{-+}$ case.
If both vector bosons are real, the fraction of longitudinally
polarized vector bosons is given by
\begin{eqnarray}
{\Gamma_L (H,A \rightarrow VV) \over \Gamma_L + \Gamma_T} &=&
\left\{ \begin{array}{lr}
\displaystyle {1 - 4 \rho + 4 \rho^2 \over 1 - 4 \rho + 12 \rho^2 } &
\mbox{for}~~H [0^{++}] \nonumber \\[5mm]
0  & \mbox{for}~~A [0^{-+}] \end{array} \right.
\label{eq:bosons_long_frac}
\end{eqnarray}
with $\rho = m_V^2/m_{H,A}^2$.
The ratio grows from 1/3 to 1 if the $0^{++}$ Higgs mass rises from
the $VV$ threshold to large values.
The $V$ helicities can be
determined by measuring the angular distributions of the
decay products in
$H, A \rightarrow VV \rightarrow (f_1 \bar{f}_2) (f_3 \bar{f}_4)$.
%

Defining two planes through the decays $V\to f_1\bar{f}_2$ and
$V\to f_3\bar{f}_4$, the distribution of the azimuthal angle
between the decay planes is given by \cite{bar94}
\begin{eqnarray}
{d \Gamma (H \rightarrow VV) \over d \phi^* }  & \sim &
1 + a_1\cos\phi^* + a_2\cos 2\phi^*
\end{eqnarray}
with
\begin{eqnarray}
a_1 & = & - { 9\pi^2 \over 32 } \:
            { \gamma_1\gamma_3(1+\beta_1\beta_3) \over
              \gamma_1^2\gamma_3^2(1+\beta_1\beta_3)^2+2 } \:
            { 2v_1a_1 \over v_1^2+a_1^2 } \:
            { 2v_3a_3 \over v_3^2+a_3^2 } \nonumber\\
a_2 & = & \frac12 \: {1 \over \gamma_1^2\gamma_3^2(1+\beta_1\beta_3)^2+2}
\end{eqnarray}
for $0^{++}$ particles and
\begin{eqnarray}
{d \Gamma (A \rightarrow VV) \over d \phi^* }  & \sim &
1 - \frac14\cos 2\phi^*
\end{eqnarray}
for $0^{-+}$ particles. As expected the $\phi^*$ asymmetry vanishes
for large scalar Higgs masses while it is independent of
the pseudoscalar Higgs mass.
For $V = W$, the charges are $v_i = a_i = 1$; for $V=Z$,
$v_i = 2I_{3i} - 4 Q_i \sin^2 \theta_W$ and $a_i = 2 I_{3i}$.
$\beta_i (\gamma_i)$ are the velocities ($\gamma$ factors)
of the vector bosons.  The angular distributions apply also
to off-shell vector bosons below threshold; the fraction
$L/(L+T)$ of longitudinally polarized vector bosons has been
discussed for this case in Ref.~\citenum{bar94}.

\vspace{1cm}

\section{Higgs--Strahlung in  $e^+ e^-$  Collisions}

The analysis of the angular distributions in the
Higgs--strahlung process
\begin{eqnarray}
e^+ e^- \rightarrow ZH
\label{eq:higgs-strahlung}
\end{eqnarray}
provides sensitive tests of the spin and parity of the
scalar Higgs particle $H$ \cite{bar94}. The explicit form
of the angular distribution can be written
\begin{eqnarray}
{ d \sigma(ZH) \over d \cos \theta} & \sim &
\beta^2 \sin^2 \theta + 8 {m_Z^2 \over s}
\label{eq:ZH_dist}
\end{eqnarray}
For high energies the $Z$ boson is produced in a state
of longitudinal polarization,
\begin{eqnarray}
{\sigma (Z_L) \over \sigma (Z_L) + \sigma (Z_T)} & = &
1 - {8 \over 12 + \beta^2 s/m_Z^2 }
\label{eq:zh_long_frac}
\end{eqnarray}
with $\beta^2 = [ 1 - (m_H + m_Z)^2/ s] [ 1 - (m_H - m_Z)^2/ s]$.
By contrast, if the produced Higgs boson were a pseudoscalar
particle, the $Z$ boson would be in a transversely polarized
state, independently of the energy, and the angular distribution
would read
$ d \sigma (ZA) / d \cos \theta = 1 - {1 \over 2} \sin^2 \theta$
instead.  We thus conclude that the angular distribution and
the fraction of Higgs--strahlung events with longitudinal $Z$
polarization provide tests which are sensitive to the spin and
parity of the scalar Higgs particles.

\vspace{1cm}

\section{ $\gamma \gamma$  Production of Higgs Particles}

The colliding $\gamma$ beam reaction
\begin{eqnarray}
\gamma \gamma & \rightarrow & H, A
\label{eq:gamma_to_higgs}
\end{eqnarray}
has long been recognized (see e.g. \citer{gh93,gg92})
as an important instrument to study the properties of Higgs
particles.  Since the $\gamma \gamma$ coupling, accessible
through the cross section\footnote{The Breit--Wigner distribution can in
general be approximated by the $\delta$-function
$\delta(s_{\gamma \gamma} - M^2)$.} of this process
\begin{eqnarray}
\sigma (\gamma \gamma \rightarrow H, A) & = &
\frac{8\pi^2}{M} \Gamma (H, A \rightarrow \gamma \gamma)
\frac{M\Gamma_{\mbox{{\scriptsize tot}}}/\pi}{(s_{\gamma\gamma}-M^2)^2+
(M\Gamma_{\mbox{{\scriptsize tot}}})^2}
\label{eq:gamma_crsect}
\end{eqnarray}
is mediated by loops of all charged particles with
non--zero mass, these
measurements can virtually reveal the action of new particle
species and help discriminate, for instance, the light Higgs
boson of supersymmetric theories from that of the Standard Model.
Using linearly polarized photon beams, the parity of
the produced Higgs boson can be measured directly \cite{gg92,bar94}.
While the polarization vectors of the two photons must be parallel
to generate $0^{++}$ scalar particles, they must be perpendicular
for $0^{-+}$ pseudoscalar particles \cite{yang49},
\begin{eqnarray}
{\cal M} (\gamma \gamma \rightarrow H [0^{++}]) & \sim &  \vec \epsilon_1
	\cdot \vec \epsilon_2 \label{eq:amp_phot-to-H} \nonumber \\
{\cal M} (\gamma \gamma \rightarrow A [0^{-+}]) & \sim & \vec \epsilon_1
	\times \vec \epsilon_2 \cdot \vec k_\gamma
\label{eq:amp_phot-to-A}
\end{eqnarray}

\vspace{3mm}

High energy colliding beams of linearly polarized photons can be
generated by Compton back-scattering of linearly polarized laser
light on electron/positron bunches of $e^+ e^-$
linear colliders \cite{ginz84}.
The linear polarization transfer from the laser photons to the
high-energy photons is described by the $\xi_3$ component of the
Stokes vector. The length of this vector depends on the final state
photon energy and on the value of the parameter
$x_0 = 4E_e\omega_0/m_{e}^{2}$
where $E_e$ and $\omega_0$ are the electron/positron and laser
energies, respectively. The requirement to operate below the $e^\pm$
pair threshold in collisions between the laser $\gamma$'s and the
high energy $\gamma$'s implies the upper bound
$x_0\: \simlt\: 4.83$; lower bounds
on $x_0$ depend on the available laser frequencies in the red/infrared
regime. As shown in Fig.\ \fcite{f_gamma}, the linear polarization
transfer is large for small values of $x_0$ if the photon energy
$y = E_\gamma/E_e$ is close to the maximum value, i.e.
\begin{eqnarray}
y \to
y_{\mbox{{\scriptsize max}}} & = & {x_0 \over 1+x_0}
\end{eqnarray}
for a given parameter $x_0$.
Adopting the usual definition
\begin{eqnarray}
r & = &  {y \over  x_0 (1 - y)} \le 1
\end{eqnarray}
[where the maximum is reached for
$y \rightarrow y_{\mbox{{\scriptsize max}}}$],
the $\gamma$ energy spectrum is given by
\begin{eqnarray}
\phi_0(y) & = & {1 \over 1 - y} + 1 - y - 4 r (1 - r)
\label{eq:phot_E_spec}
\end{eqnarray}
and the Stokes component $\xi_3 (y) = \phi_3(y) / \phi_0(y)$
follows from
\begin{eqnarray}
\phi_3(y) & = & 2 r^2
\end{eqnarray}
The maximum value of the Stokes vector $\xi_3(y)$ is reached for
$y = y_{\mbox{{\scriptsize max}}}$,
\begin{eqnarray}
\xi_{3}^{\mbox{{\scriptsize max}}} & = &
{2(1+x_0) \over 1+(1+x_0)^2}
\end{eqnarray}
and approaches unity for small values of $x_0$.

\vspace{3mm}

The overall $\gamma \gamma$ luminosity is assumed to be
of the same size as the $e^+ e^-$ collider luminosity.

\vspace{3mm}

Since only part of the laser polarization is transferred to the
high-energy photon beam, it is useful to define the polarization
asymmetry ${\cal A}$ as
\begin{eqnarray}
{\cal A} & = & {N^\parallel - N^\perp \over N^\parallel + N^\perp}
\label{eq:phot_asy_def}
\end{eqnarray}
where $N^\parallel$ and $N^\perp$ denote the number of
$\gamma \gamma$ events with the initial laser polarizations
being parallel and perpendicular, respectively. It follows
from eqn.(\ref{eq:phot_asy_def}) that
\begin{eqnarray}
{\cal A} (\gamma \gamma \rightarrow H [0^{++}]) & = &
+ {\cal A}\nonumber\\
{\cal A} (\gamma \gamma \rightarrow A [0^{-+}]) & = &
- {\cal A}
\label{eq:phot_asy_rels}
\end{eqnarray}

\vspace{3mm}

For resonance production, the asymmetry
${\cal A}$ can be expressed by the
appropriate luminosity factors,
\begin{eqnarray}
{\cal A} &=& { <\phi_3 \phi_3 >_\tau \over < \phi_0 \phi_0 >_\tau }
\label{eq:phot_asy}
\end{eqnarray}
with
\begin{eqnarray}
<\phi_i \phi_i>_\tau & = & {1 \over N^2}
	\int_{ \tau / y_{\mbox{{\tiny max}}}}
         ^{y_{\mbox{{\tiny max}}}}
	{dy \over y} \phi_i(y) \phi_i(\tau /y)
\label{eq:phi_prod_def} \nonumber
\end{eqnarray}
The Drell--Yan parameter is given by
\begin{equation}
\tau \: = \: M^2/s_{e^+ e^-}
\end{equation}
where $M$ is the resonance mass and
$s_{e^+e^-}$ is the $e^\pm$ collider energy-squared.
The maximum value of $\tau$ is bounded by the energy transfer
\begin{eqnarray}
\tau_{\mbox{{\scriptsize max}}} & = &
y^{2}_{\mbox{{\scriptsize max}}} \nonumber \\
& = & \left [{x_0 \over 1+x_0}\right ]^2
\end{eqnarray}
The normalization factor $N$ is defined as
\begin{eqnarray}
N & = & \int_0^{y_{\mbox{{\tiny max}}}} dy \phi_0(y)
\label{eq:phi_prod_moredef} \nonumber\\
& = & \ln (1+x_0)\left [1-{4 \over x_0} - {8 \over x_0^2}\right ]
      + {1 \over 2} + {8 \over x_0} - {1 \over 2(1+x_0)^2}
\end{eqnarray}
The components
$<\phi_i \phi_i>_\tau$ can easily be calculated, with the
somewhat cumbersome results
\begin{eqnarray}
<\phi_3 \phi_3>_\tau & = & {1 \over N^2}{4 \tau^2 \over x_0^4}
\left\{2\ln\left[{\sqrt{\tau} \over x_0 - \tau(1 + x_0)} \right]
\displaystyle {\tau + 1 \over (\tau - 1)^3}
+ \displaystyle {2 \over (\tau - 1)^2}
{ \tau (1 + x_0)^2 - x_0^2 \over \tau - x_0 + x_0 \tau } \right\}
\nonumber \label{eq:phi_comp_3}
\end{eqnarray}
and
\newpage
\begin{eqnarray}
<\phi_0 \phi_0>_\tau & = & {1 \over N^2}
\left\{ { 4 \over x_0^4 (1 + x_0) (\tau - 1)^2 }
        { \tau(1 + x_0)^2 - x_0^2 \over	\tau - x_0 + x_0 \tau }
\bigg [ x_0^3 (\tau - 1)^2 (\tau - x_0 + x_0 \tau )
\right. \nonumber \\
&& - 2 x_0 (1 + x_0) (\tau - 1) ( x_0 (\tau^2 - 2 \tau + 2) - 4 \tau)
   + 8 \tau^2 (1 + x_0) \bigg ] \nonumber \\
&& + \ln \left[ {\sqrt{\tau} \over x_0 - \tau(1 + x_0)} \right]
	{ 8 \over x_0^4 (\tau - 1)^2 }
\left[ 2 x_0^2 (2 \tau^2 - 4 \tau + 1)
 \vphantom{\tau^2 \over (\tau - 1)}
   + 8 x_0 \tau^2 \right.  \nonumber \\
&& \left.
   - x_0^3 (\tau -1) (\tau^2 - \tau + 2)
   + {x_0^4 \over 4} (\tau - 1) ( \tau^2 - 2 \tau + 2)
   + {4\tau^2 \over (\tau - 1)} (\tau + 1)  \right] \nonumber \\
&& \left.
   + \ln \left[ {x_0 \over \sqrt{\tau} (1 + x_0) }\right] {4 \over x_0^2}
   \left[ 2 x_0 (\tau + 2) + x_0^2 + 4 \right]
	\vphantom{4 x_0^2 \over x_0^4} \right\}
\label{eq:phi_comp_0}
\end{eqnarray}

The correlation functions $<\phi_i \phi_i>_\tau$ and
the asymmetry ${\cal A}$
are displayed in Figs.~\fcite{f_phot_comp} and \fcite{f_phot_asy}.
As expected, the maximum sensitivity
\begin{eqnarray}
{\cal A}_{\mbox{{\scriptsize max}}} & = &
(\xi_{3}^{\mbox{{\scriptsize max}}})^2 \nonumber \\ & = &
\left [ {2(1 + x_0) \over 1 + (1 + x_0)^2} \right ]^2
\label{eq:asym_max}
\end{eqnarray}
is reached for small values of $x_0$ and near the upper bound of
$\tau\: \simlt\: \tau_{\mbox{{\scriptsize max}}} (x_0)$,
i.\ e.\ if the energy is just sufficient to produce
the Higgs particles in the $\gamma\gamma$ collisions.\footnote{Since the
luminosity vanishes at $\tau = \tau_{\mbox{{\scriptsize max}}}$, one will
set the operating conditions
in practice such that $\tau \simlt \tau_{\mbox{{\scriptsize max}}}$
allows for a sufficiently large luminosity $< \phi_0 \phi_0 >_\tau$, as
illustrated in Figs.~\fcite{f_crsect_hsm} to \fcite{f_crsect_sha}.}
Typical energies for electron/positron and laser beams are shown in
Table~1 for a sample of $x_0$ values corresponding to large and small
asymmetries ${\cal A}_{\mbox{{\scriptsize max}}}$.
\begin{table}[h,t]
\begin{center}
\renewcommand{\arraystretch}{1.3}
\begin{tabular}{|c|cc|c|cc|} \hline
 $x_0$ & ${\cal A}_{\mbox{{\scriptsize max}}}$ &
 $ \sqrt{\tau_{\mbox{{\scriptsize max}}}}$ &
 $M_H$~[GeV] & $E_e$~[GeV] & $\omega_0$~[eV] \\
 & & & & (at $\sqrt{\tau_{\mbox{{\scriptsize max}}}}$) &  \\ \hline\hline
  0.5  & 0.85 & 0.33 & 100 & 150 & 0.22 \\
       &      &      & 200 & 300 & 0.11 \\
       &      &      & 300 & 450 & 0.07 \\ \hline
  1.0  & 0.64 & 0.5  & 100 & 100 & 0.65 \\
       &      &      & 200 & 200 & 0.33 \\
       &      &      & 300 & 300 & 0.22 \\ \hline
  2.0  & 0.36 & 0.67 & 100 & 75  & 1.74 \\
       &      &      & 200 & 150 & 0.87 \\
       &      &      & 300 & 225 & 0.58 \\ \hline
  4.83 & 0.11 & 0.83 & 100 & 60.4& 5.22 \\
       &      &      & 200 & 121 & 2.61 \\
       &      &      & 300 & 181 & 1.74 \\ \hline
\end{tabular}
\end{center}
\caption[dummy]{{\em Electron/positron energies
         $E_e$ of the $e^+e^-$ linear collider and laser
         $\gamma$ energies
         $\omega_0$ for a sample of Higgs masses if the parameter
         $x_0 = 4E_e\omega_0/m_e^2, which $ determines the maximum
         asymmetry ${\cal A}_{\mbox{{\scriptsize max}}}$
         at $\sqrt{\tau_{\mbox{{\scriptsize max}}}} =
         y_{\mbox{{\scriptsize max}}}$, is varied from
	 small to large values.}}
\end{table}
\vspace{3mm}

The measurement of the Higgs parity in $\gamma\gamma$ collisions
will be a unique method in areas of the parameter space where
the Higgs coupling to heavy $W,Z$ bosons are small and the top
quark decay channels are closed so that the Higgs particles
decay preferentially to $b,c,\tau$ fermions. This method is
therefore most useful for particles in extended Higgs models
while the profile of the $\cal{SM}$ Higgs particle can be
determined in Higgs decay and $e^+e^-$ production processes.

It must therefore be shown that the
background events  from heavy quark production, $\gamma \gamma
\rightarrow b \bar{b}$ and $c \bar{c}$, can be suppressed sufficiently well.
This is a difficult task \cite{gh93,halzen93} for $b$ quarks.
For charm quarks the situation looks superficially even more miserable
since the $c \bar{c}$ production cross sections are $\sim 16/4/1$ times
larger than the $b \bar{b}$ cross sections for
direct/1--resolved/2--resolved $\gamma$ events at high energies.
However,
to simplify the discussion we shall assume, for the time
being, $\mu$-vertexing
to be perfect and we neglect the contamination from $c \bar{c}$
production. [This is actually not far from reality, as discussed
 in Ref.\ \citenum{bor93}.]

\vspace{3mm}

Three components contribute to the $b \bar{b}$ background events:
direct $\gamma \gamma$ production, the 1--resolved photon process
$\gamma\gamma(\to g) \to b\bar{b}$, and the 2--resolved photon
process $\gamma\gamma(\to gg) \to b\bar{b}$ \cite{gh93,halzen93,drees}.

\vspace{3mm}

The cross section for direct $\gamma \gamma$ production
\begin{eqnarray}
\gamma \gamma & \rightarrow & b \bar{b}
\label{eq:dir_prod_b}
\end{eqnarray}
can be easily calculated at the tree level for linearly-polarized
photons. Effects due to higher-order QCD corrections have been shown
to be modest in the unpolarized case \cite{drees} and, hence, can be
safely neglected for asymmetries. The cross sections are given by
\begin{eqnarray}\label{35}
{ d \sigma^\parallel \over dy d \phi} & = &
{6 \alpha^2 Q_{b}^{4}  \over s}\:
{ 1 + e^{-4y} + 2 e^{-2y} \sin 4 \phi \over (1 + e^{-2y})^2 }
\label{eq:b_crsect_para} \nonumber \\
{ d \sigma^\perp \over dy d \phi} & = &
{6 \alpha^2 Q_{b}^{4}  \over s}\:
{ 1 + e^{-4y} - 2 e^{-2y} \sin 4 \phi \over (1 + e^{-2y})^2 }
\label{eq:b_crsect_perp}
\end{eqnarray}
for parallel and perpendicular photon polarizations, respectively,
at energies sufficiently above the quark threshold.
Note that $y$ denotes here the
rapidity of the $b$ quark and $\phi$ the azimuthal angle of the
production plane with respect to either of the photon polarization
vectors in the center-of-mass frame. If only one of the photon beams
is linearly polarized and the second beam unpolarized, the cross
section does not depend on the azimuthal angle $\phi$
for energies sufficiently above the $b\bar{b}$ threshold.
The unpolarized cross section is the average of the parallel and
perpendicular cross sections. 
As evident from eqs.(\ref{35}), the background process
$\gamma \gamma \rightarrow b \bar{b}$ does not affect the numerator
of the asymmetry ${\cal A}$ if the azimuthal angle is integrated over,
yet it does increase the denominator, i.e.
$ N^\parallel + N^\perp $ $\rightarrow$
$ [N^\parallel + N^\perp]_{\mbox{{\scriptsize Res}}} $ $+$
$ [N^\parallel + N^\perp]_{\mbox{{\scriptsize bkgd}}} $,
thus diluting the asymmetry in general by a significant amount.

\vspace{3mm}

If one or two photons are resolved into quark-plus-gluon showers,
the subprocesses $\gamma g \rightarrow b \bar{b}$ and $gg \rightarrow
b \bar{b}$ generate $b$-quark final states which are accompanied by
hadron jets spraying into in the photon direction(s).  Since gluons are
generated only in the double-splitting process $\gamma \rightarrow q
\rightarrow g$, the gluon spectrum in the photon falls off steeply
with the gluon momentum.  Therefore, the 1--and 2--resolved photon
processes are strongly suppressed if nearly all the photon energy is
needed to generate the $b \bar{b}$ final state energy.  This is
however the situation we encounter when the asymmetry ${\cal A}$
in Higgs production will be measured for $\tau$ values close to
$\tau_{\mbox{{\scriptsize max}}}(x_0)$,
cf.\ Fig.~\fcite{f_phot_asy}.
The background from 1--resolved processes is therefore expected to be
small in this kinematical configuration and negligible for the
2--resolved photon processes. [A detailed example will be presented in
Fig.~\fcite{f_crsect_hsm}(a).]
The resolved photon mechanisms therefore do not pose background problems
in our case.

\vspace{3mm}

While the signal events $\gamma \gamma \rightarrow H, A \to b\bar{b}$ are
distributed isotropically in their center-of-mass frame, the
background events are strongly peaked at zero polar angles.
This can be exploited to reject the background events by demanding
small values ($| y_{b, \bar{b}}| \simlt 1$) for the rapidities of the
$b$ and $\bar{b}$ quarks. In Figs.~\fcite{f_crsect_hsm} to
\fcite{f_crsect_sha}(a)
we compare the (unpolarized)
signal cross sections for $H_{SM}$ in the Standard Model and
$h^0/H^0/A^0$ in the minimal supersymmetric model at $\tan \beta= 1.5$
and $30$ with the background $b \bar{b}$ channels integrated over
rapidities $| y_{b, \bar{b}}| \le 1$. We have assumed a mass resolution
$\Delta M = 5~\mbox{GeV}$.\footnote{Coulomb--gluon exchange between the
top--antitop quarks mediating the $\gamma\gamma$ coupling to the
$A^0$ Higgs boson, leads to pseudoscalar S-wave [$t\bar{t}\,$]
resonance contributions near the threshold, and the lowest--order
analysis cannot be applied any more in this range,
Ref.~\citenum{abdel}.}
In these illustrative examples $x_0$ has been
chosen = 1 and the $\tau$ values have been set slightly below
$\tau_{\mbox{{\scriptsize max}}}$ to allow for sufficiently large
luminosities at the resonance $\gamma\gamma$ energy.
It is clear from these figures that the
measurement of the Higgs parity, in particular for the heavy particles,
requires high $\gamma \gamma$ luminosities. The background events
reduce the asymmetry ${\cal A}$ by a suppression factor $1/[1 + B/S]$
where $S$ ($B$) denote the number of signal (background) events.
The asymmetries including background events are displayed in
Figs.~\fcite{f_crsect_hsm} to
\fcite{f_crsect_sha}(b) for the
Standard Model Higgs particle $H_{SM}$ and for the $h^0/H^0/A^0$
particles in the minimal supersymmetric model
for $\tau$ values slightly below $\tau_{\mbox{{\scriptsize max}}}$
where the polarization asymmetry is maximal.
If an integrated luminosity of $\int {\cal L} = 100 fb^{-1}$ can
be accumulated within a few years, asymmetries down to $\sim 3 \%$
will be accessible, covering
major parts of the intermediate Higgs mass range in the Standard Model
and of the supersymmetric parameter space.
To carry out these experiments, laser operations in the red/infrared
regime are necessary in order to generate sufficiently high asymmetries.

\vspace{3mm}

Under these conditions the polarization asymmetry of the
$\cal{SM}$ Higgs particle
$H_{SM}$  can be measured in $\gamma\gamma$ collisions
throughout the relevant mass range below
$\sim 150$~GeV in the $b\bar{b}$ channel;
above this mass value Higgs decays to $Z$ bosons
(virtual/real) can be exploited to determine spin and parity.
The light scalar $\cal{MSSM}$ Higgs boson $h^0$ can be probed
in a similarly comprehensive way, except presumably for the low
mass range at large $\tan \beta$.  By contrast, the measurement
of the polarization asymmetry of the heavy scalar $\cal{MSSM}$ Higgs
particle $H^0$ appears to be very difficult in the $b\bar{b}$ channel.
In this case decays to gauge bosons and top quark pairs, as well as the
angular distribution in the Higgs-strahlung production process are
more promising, as evident from Fig.~\fcite{f_higgsdec}.
Finally, the $\gamma \gamma$ polarization measurement of the parity
in the very interesting case of the pseudoscalar $A^0$ Higgs particle
appears feasible throughout most of the parameter range below the
top threshold; $A^0 \rightarrow t \bar{t}$ decays can be exploited
for masses above this threshold.

These figures represent illustrative examples.  Optimizing
the run parameters [$x_0/y/\Delta y_{b \bar{b}}$] will help refine the
analysis if necessary.

\vspace{1cm}

\section{Conclusions}
The analyses in the preceding sections provide a transparent picture
of prospects to determine the external quantum numbers
$J^{PC} = 0^{++}$ and $ 0^{-+}$ of the scalar and pseudoscalar Higgs
particles experimentally.

\vspace{3mm}

\noindent
(i) For scalar $0^{++}$ Higgs particles the measurement is easy if the
particles couple strongly enough to the $W,Z$ gauge bosons. In
this case Higgs decays to vector boson pairs can be exploited at the
LHC and $e^+e^-$ linear colliders if the Higgs mass is sufficiently large.
In the intermediate mass range, the production through Higgs--strahlung
off the (virtual/real) $Z$ boson in $e^+e^-$ collisions provides a
complementary method. The angular distributions of the decay and
production processes are characteristic for the spin and parity of
the Higgs particles. This has been demonstrated by confronting the
$0^{++}$ predictions with distributions derived for other assignments
of these quantum numbers. While the $\cal{SM}$ Higgs sector can be covered
completely this way, the method can be applied only in a restricted range
of the $\cal{SUSY}$ Higgs parameter space.

\vspace{3mm}

\noindent
(ii) Since the pseudoscalar $A^0$ Higgs boson in supersymmetric
extensions of the Standard Model does not couple directly
to gauge bosons, other methods must be used to measure its parity.
Above the $t\bar{t}$ decay threshold, correlations among the $(bW)$
decay products of the $t$ and $\bar{t}$ quarks allow us to determine
the negative parity in a straightforward way.  Analogous information is
provided by the hadronic decays of $\tau$-pairs produced in Higgs decays
below the top pair production threshold.

\vspace{3mm}

\noindent
(iii) In major areas of the $\cal{SUSY}$ parameter space the fusion of Higgs
particles in linearly polarized $\gamma\gamma$ collisions provides
an alternative method to these classical tools. Parallel / perpendicular
$\gamma\gamma$ beams generate $0^{++}$ / $0^{-+}$ Higgs particles,
respectively. By choosing the frequency of the laser light in the
Compton back--scattering process for generating high energy $\gamma$
beams as low as possible and the original $e^\pm$ energy just high
enough to produce the Higgs particles in the $\gamma\gamma$ collisions,
one can find an optimum point for operating the experiment at which
the asymmetry is maximal and the Higgs production rate sufficiently
large. This is a particularly interesting method for measuring the
negative parity of $A^0$ Higgs bosons in $\cal{SUSY}$ extensions
of the Standard Model which is hard to determine otherwise.
{\em In summa} Besides the measurement of the Higgs
$\gamma\gamma$ width, the prospects of determining the external
quantum numbers of the Higgs particles render dedicated
$\gamma\gamma$ collisions a very interesting instrumental
option for future $e^+e^-$ linear colliders.

\newpage

\vspace{1cm}
{\noindent\large\bf Acknowledgements}
\vspace{3mm}

\noindent The authors would like to thank K.~Hagiwara for
helpful discussions.
This work is supported in part by BMFT under contract 056KA93P,
by KBN under grant 2P30225206 and by
Deutsche For\-schungs\-ge\-mein\-schaft DFG. MK gratefully
acknowledges the hospitality extended to him by the DESY Theory Group.

\newpage

\begin{thefiglist}{99}
\figitem{f_higgsdec}Branching ratios of Higgs decay modes
in the Standard Model and its minimal supersymmetric
extension. For the sake of clarity only those branching ratios
are displayed which are of interest for the subsequent analyses.
The top mass has been set to $m_t = 150$~GeV.
\figitem{f_tau_angdist} Definition of the pion angles $\theta_\pm^*$ and
$\phi^*$ in the decays $\tau^{\pm} \to \pi^{\pm}\stackrel{(-)}{\nu_{\tau}}$.
The polar angles $\theta_\pm^*$ are defined in the $\tau_\pm$
rest frames, with respect to the $\tau^-$ direction.  $\phi^*$ remains
unchanged under boosts along the $\tau^+,\tau^-$ axis.
The angle $\delta$ separates the pions in the Higgs rest frame; it
is close to $\pi$ because of the large Lorentz boosts from $\tau$ to
Higgs rest frames.
\figitem{f_pidist} Distributions of the decay $H,A \rightarrow \tau^+ \tau^-
\rightarrow \pi^+ \overline{\nu}_{\tau} \pi^- \nu_{\tau}$
in the angle between the
pions as measured in the Higgs rest frame.  The plot is for non-zero
pion mass and for $m_{H,A} = 60, 150$ GeV.  The distributions for
scalar (pseudoscalar) Higgs particles are drawn with solid (dashed) lines.
\figitem{f_rhodist} Distributions for $H, A \rightarrow \tau^+ \tau^-
\rightarrow \rho^+ \overline{\nu}_{\tau} \rho^- \nu_{\tau}$ in the angle
between the $\rho$'s.  The difference between scalar (solid)
and pseudoscalar (dashed) distributions is much smaller than in the
pion case.  This effect is due mainly to the suppression factor
$ (m_\tau^2 - 2 m_\rho^2)^2/(m_\tau^2 + 2 m_\rho^2)^2$ discussed in
the text.
\figitem{f_w_angdist}Definitions of the angles $\theta_\pm^*$ and
$\phi^*$ in the decay $H\to t\bar{t}\to (bW^+)+(\bar{b}W^-)$. The polar
angles $\theta_\pm^*$ are defined in the $t$ and $\bar{t}$ rest frames,
with respect to the $t$ flight direction.
The azimuthal angle $\phi^*$ remains
unchanged under boosts along the $t,\bar{t}$ axis.
\figitem{f_leptdist} Distributions of the decays $H,A \rightarrow
t \bar{t} \rightarrow (b l^+ \nu) (\bar{b} l^- \overline{\nu})$ in the angle
between the charged leptons.  For massless $b$'s and leptons, the only
remaining dependence is on the ratio of top to Higgs masses.  Scalar
(solid) and pseudoscalar (dashed) distributions are shown for
$m_t = 150~\mbox{GeV}$ and $m_{H,A} = 400, 1000~\mbox{GeV}$.
\figitem{f_gamma}The Stokes component $\xi_3(y)$, determining the
high energy $\gamma$ polarization, for various values of
$x_0 = 4E_e\omega_0/m_{e}^{2}$ [$E_e = e^\pm$ colliding beam energy;
$\omega_0$ = laser energy].
\figitem{f_phot_comp}The luminosity components $<\phi_3 \phi_3>_\tau$
and $<\phi_0 \phi_0>_\tau$ building up the linear polarization
asymmetry ${\cal A}$.
\figitem{f_phot_asy}The polarization
asymmetry ${\cal A}$ for various values of $x_0$.
\figitem{f_crsect_hsm} Standard Model Higgs particle $H_{SM}$:
signal and background cross sections for $b\bar{b}$ final states (a),
and the polarization asymmetry ${\cal A}$ including the background
process (b). The top mass has been chosen $m_t = 150$~GeV in this and the
subsequent figures.
\figitem{f_crsect_sh1}$\cal{MSSM}$ Higgs particle $h^0$:
signal and background cross sections for $b\bar{b}$ final states (a),
and the polarization asymmetry ${\cal A}$ including the background
process (b).
\figitem{f_crsect_sh2} $\cal{MSSM}$ Higgs particle $H^0$:
signal and background cross sections for $b\bar{b}$ final states (a),
and the polarization asymmetry ${\cal A}$ including the background
process (b).
\figitem{f_crsect_sha} $\cal{MSSM}$ Higgs particle $A^0$:
signal and background cross sections for $b\bar{b}$ final states (a),
and the polarization asymmetry ${\cal A}$ including the background
process (b).
\end{thefiglist}

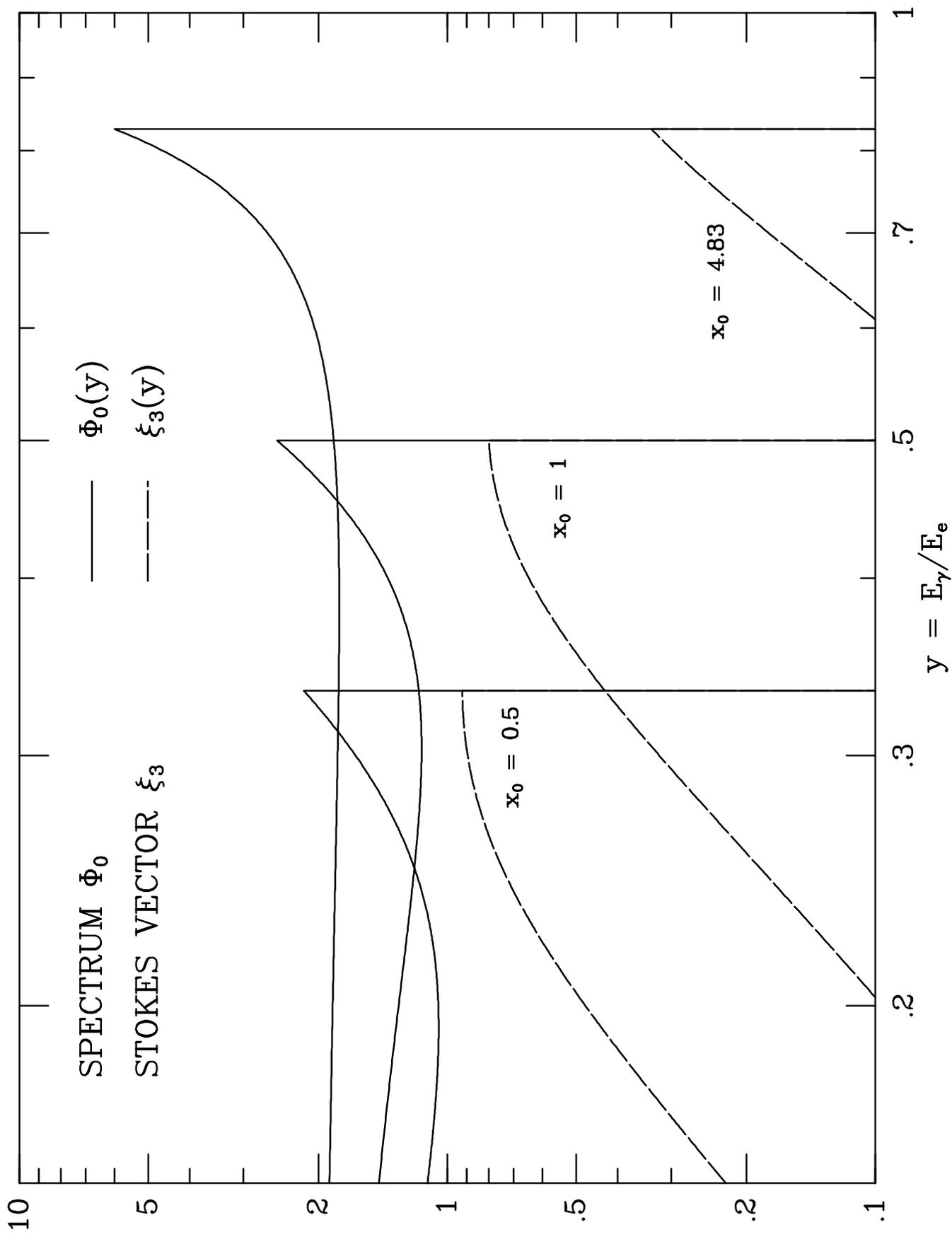
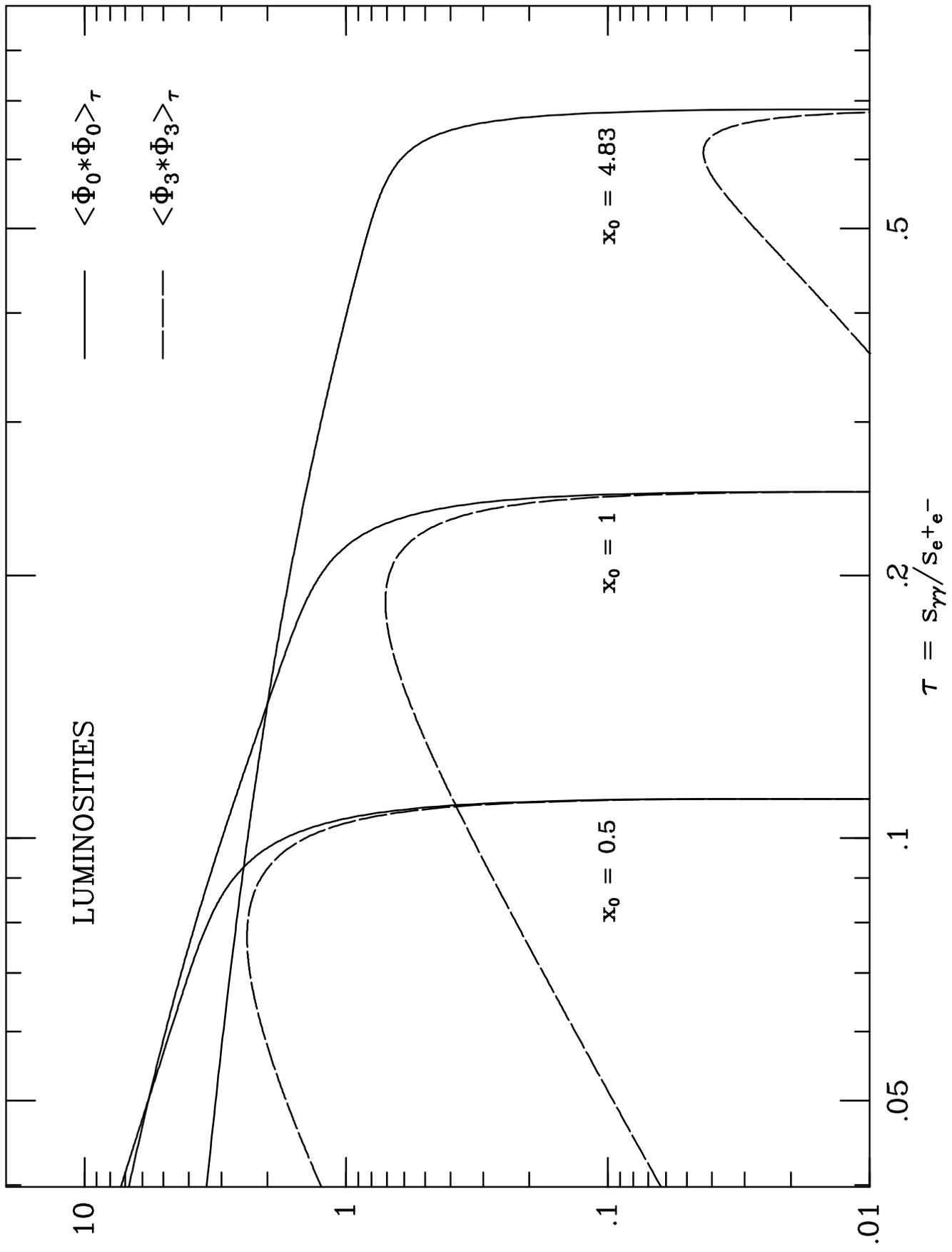
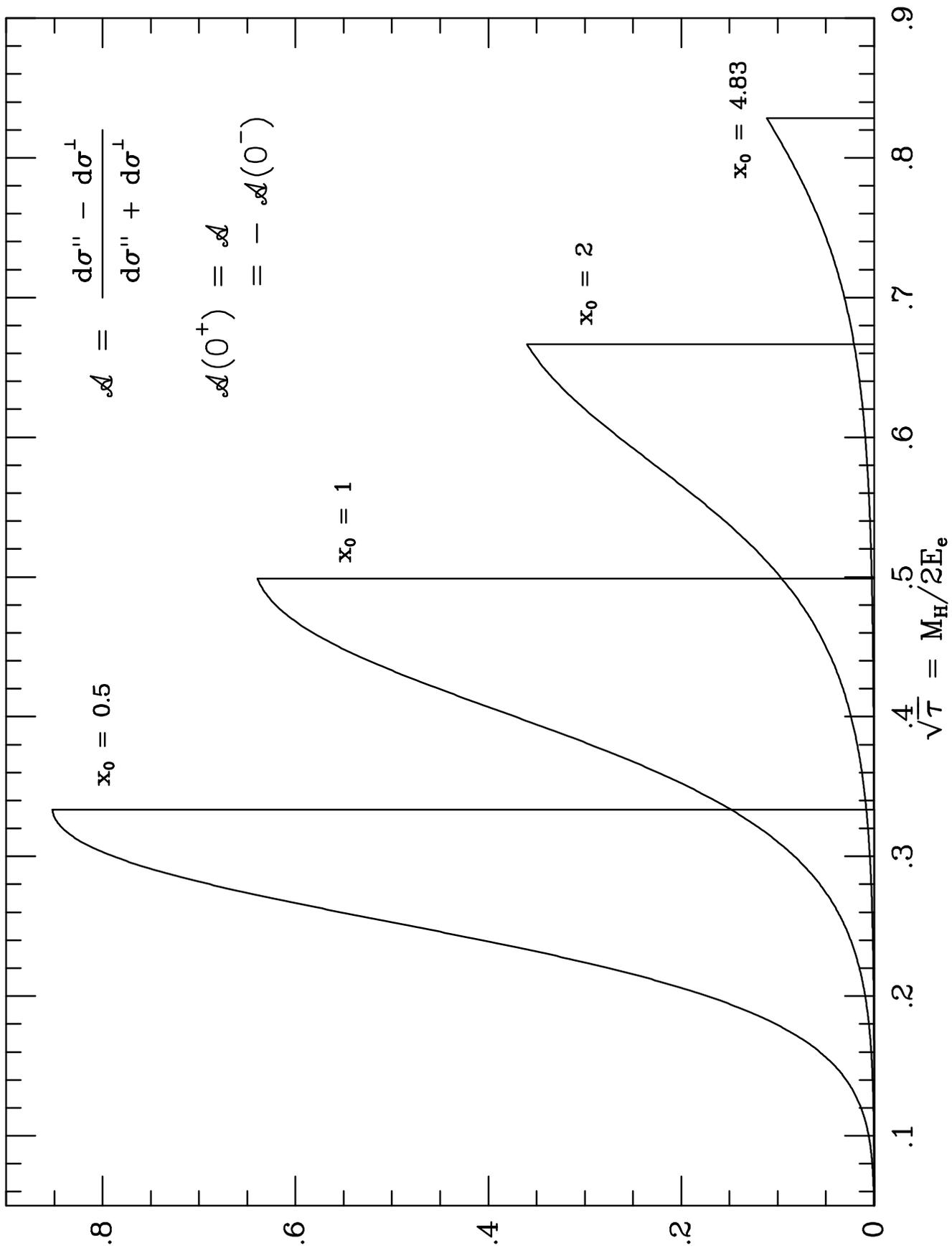
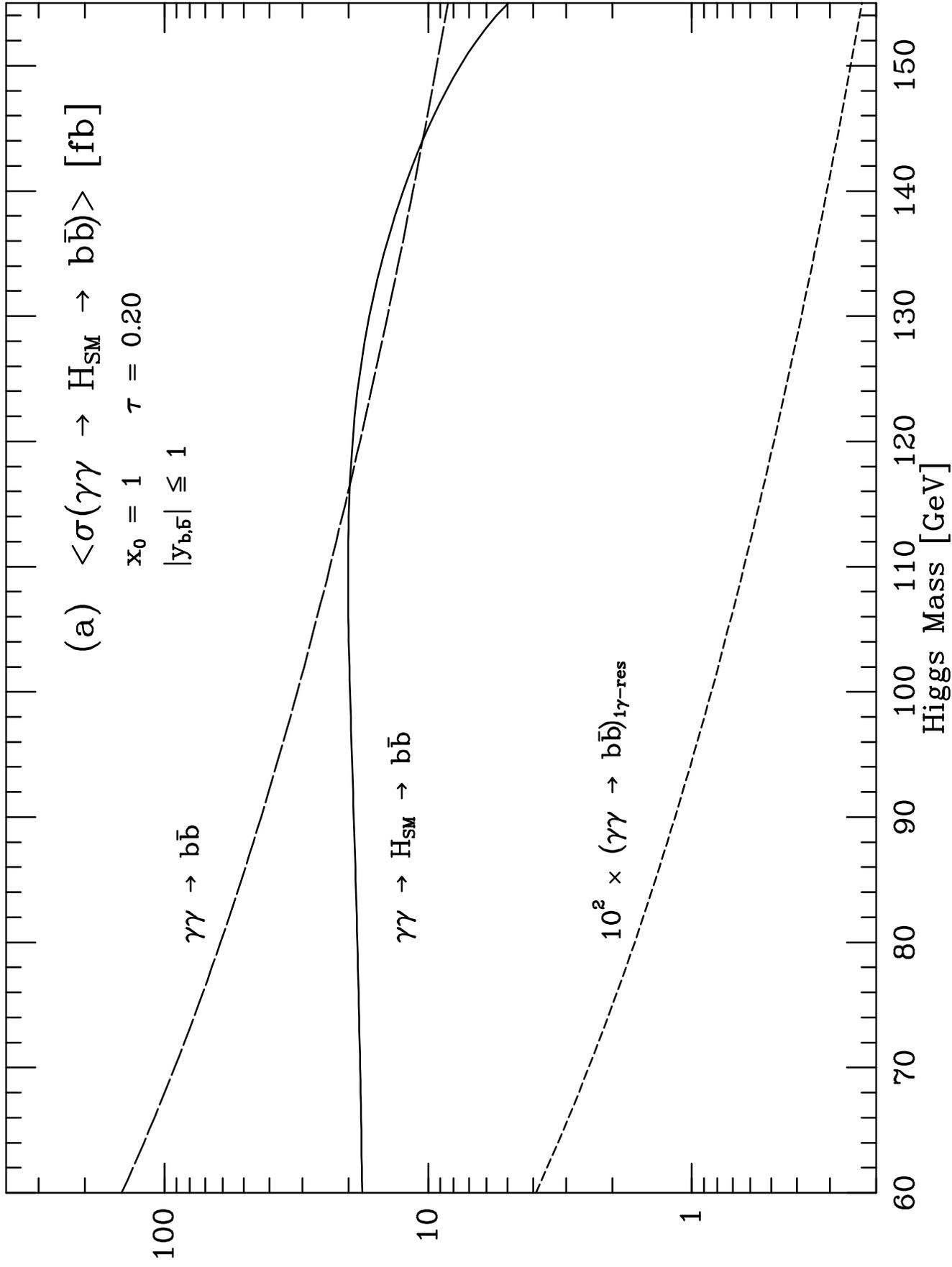
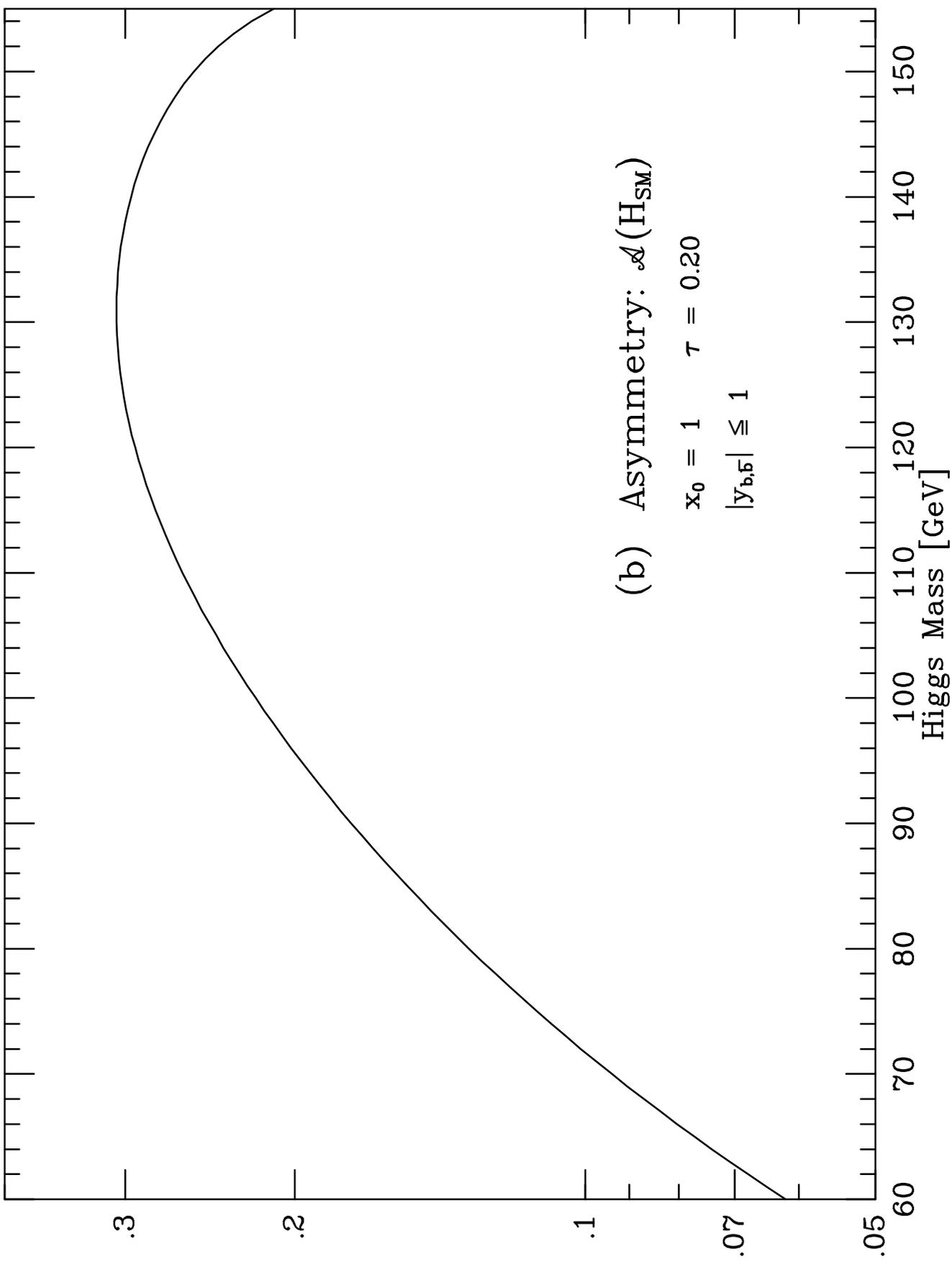
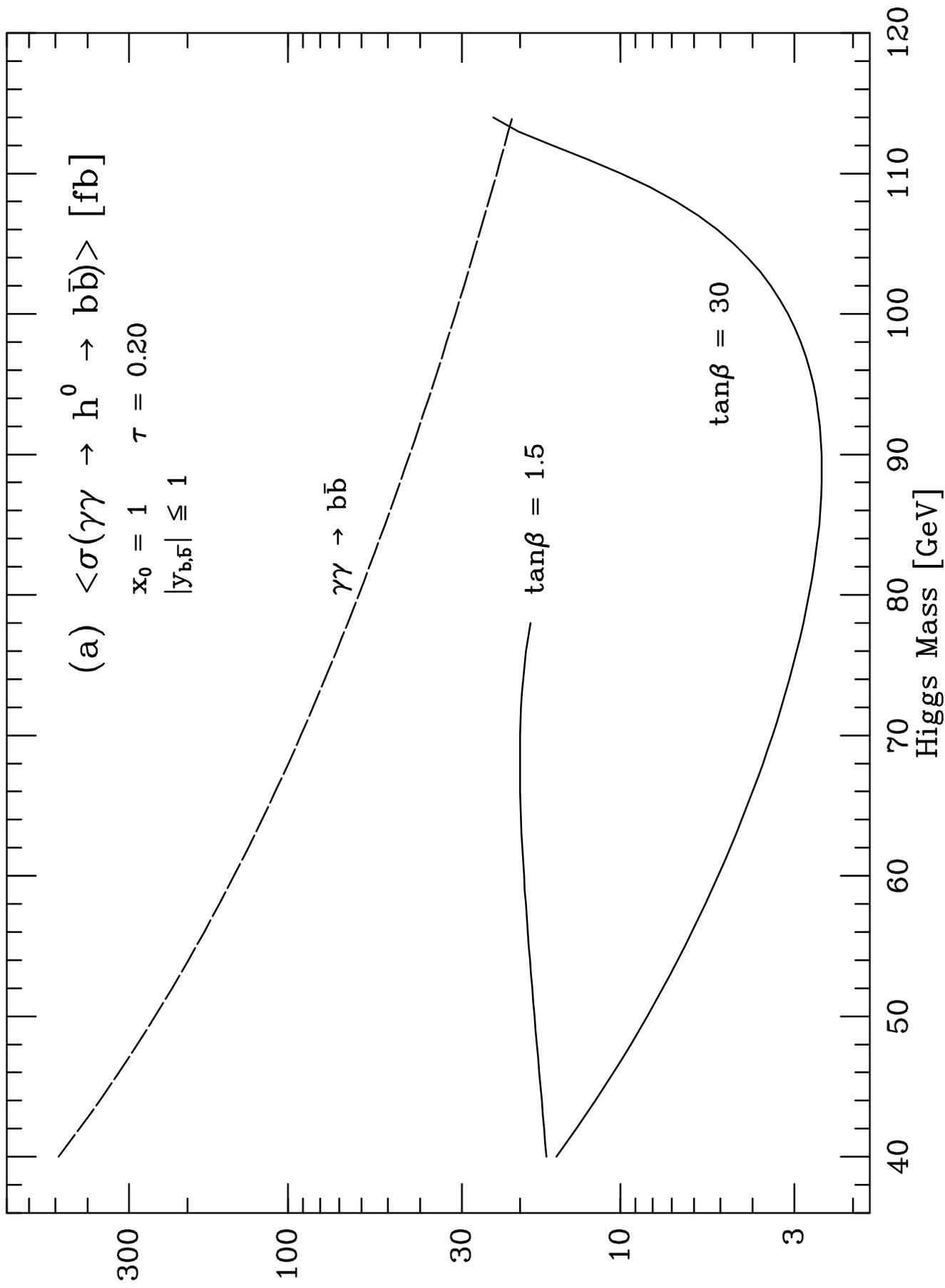
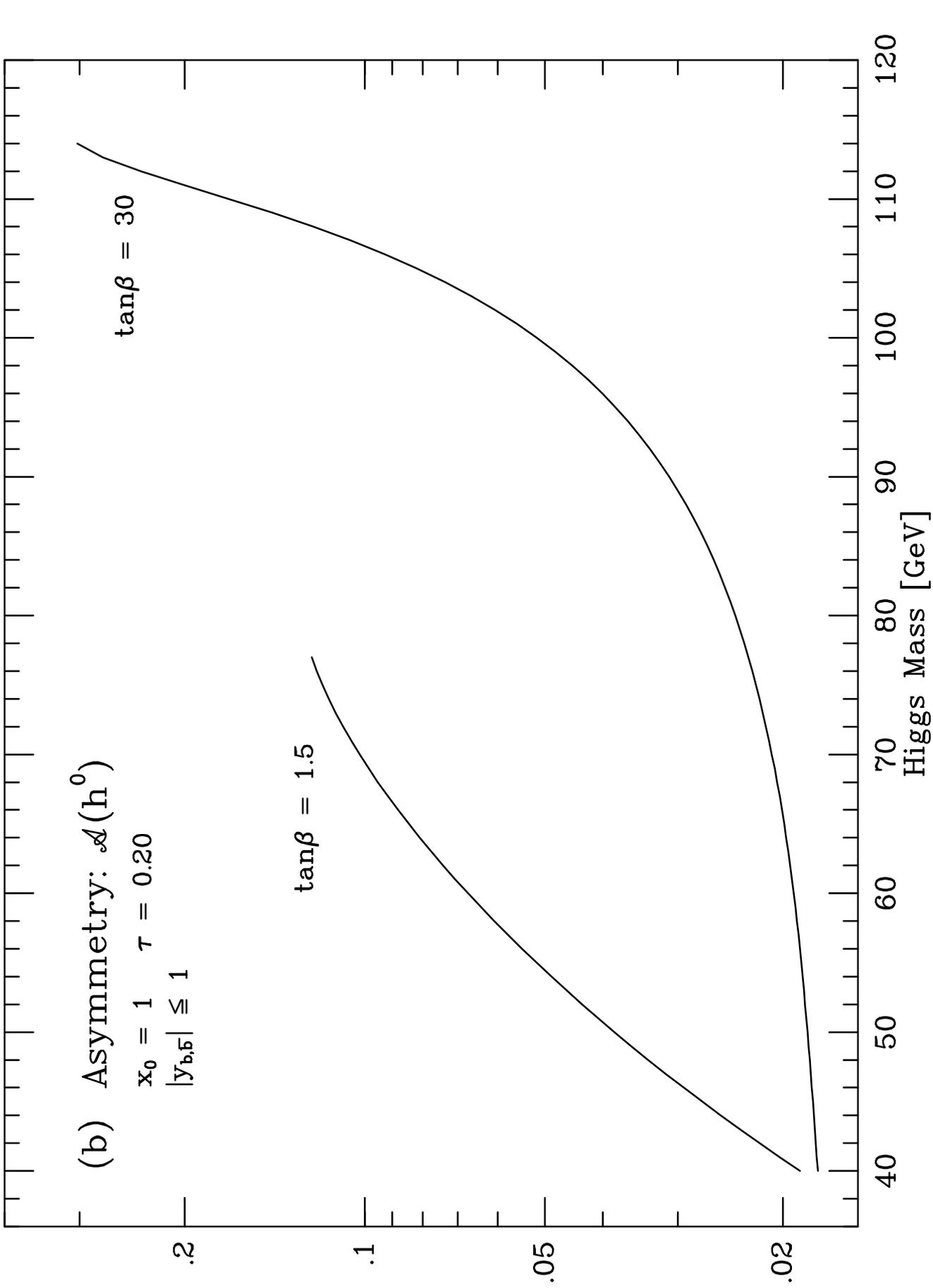
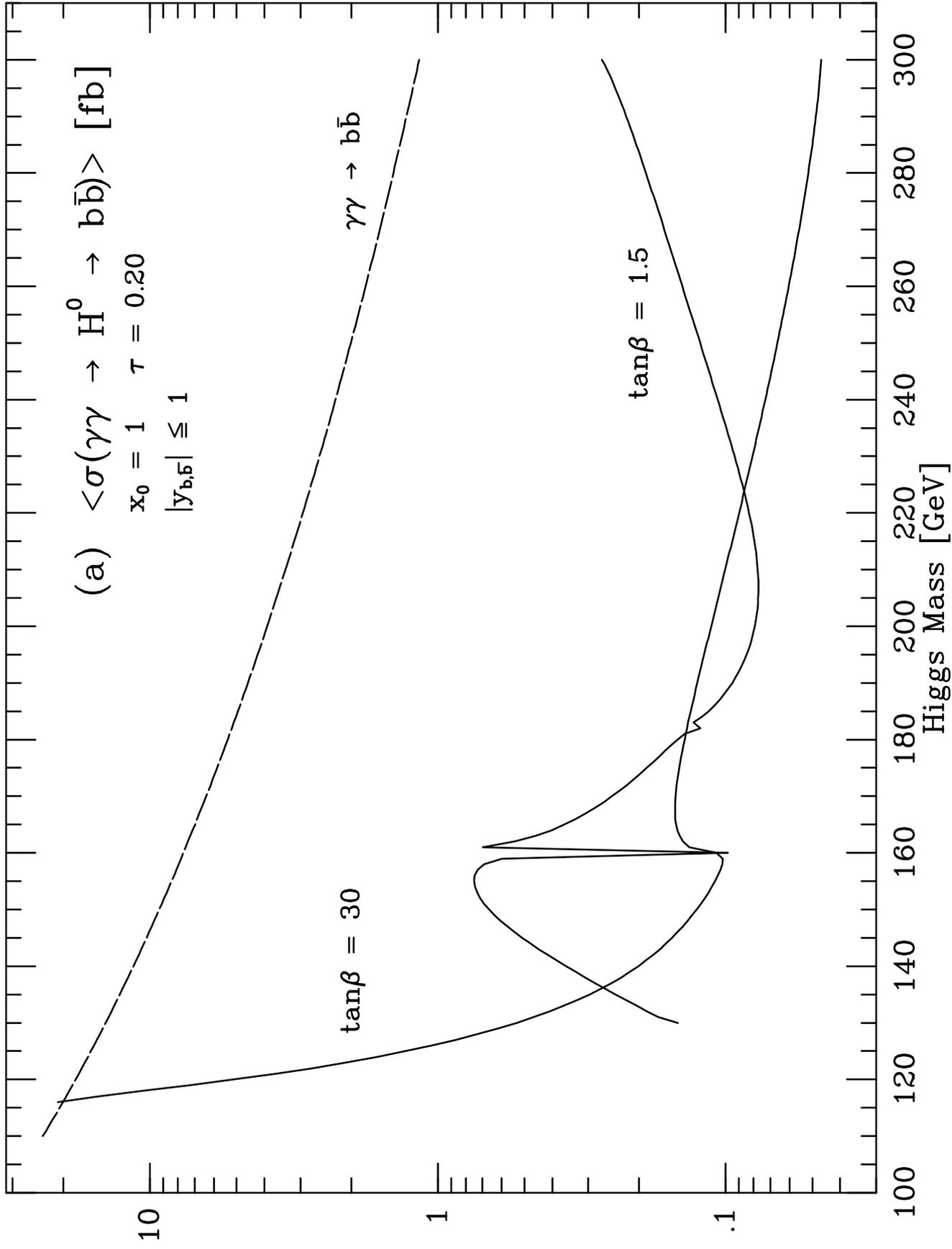
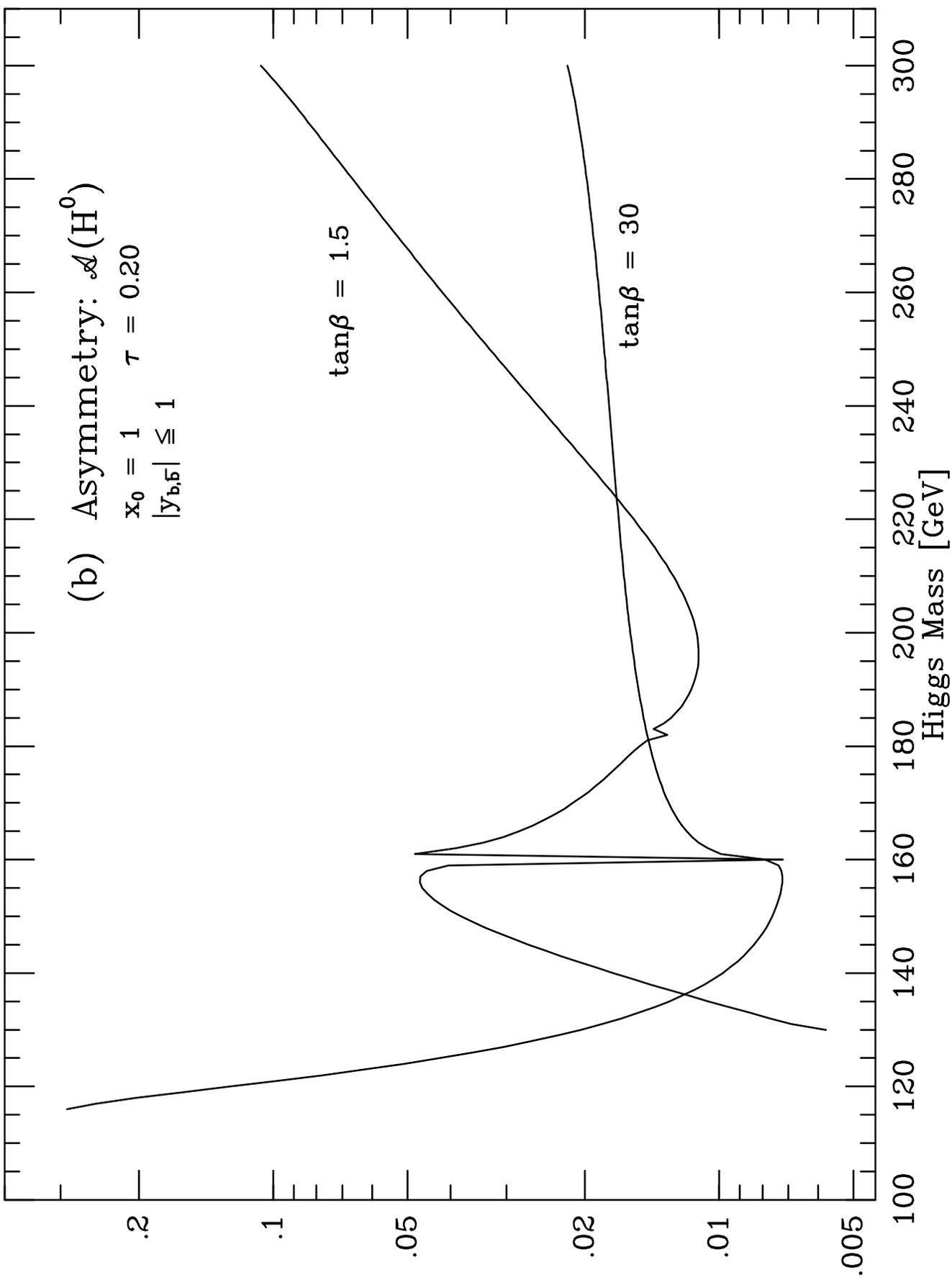
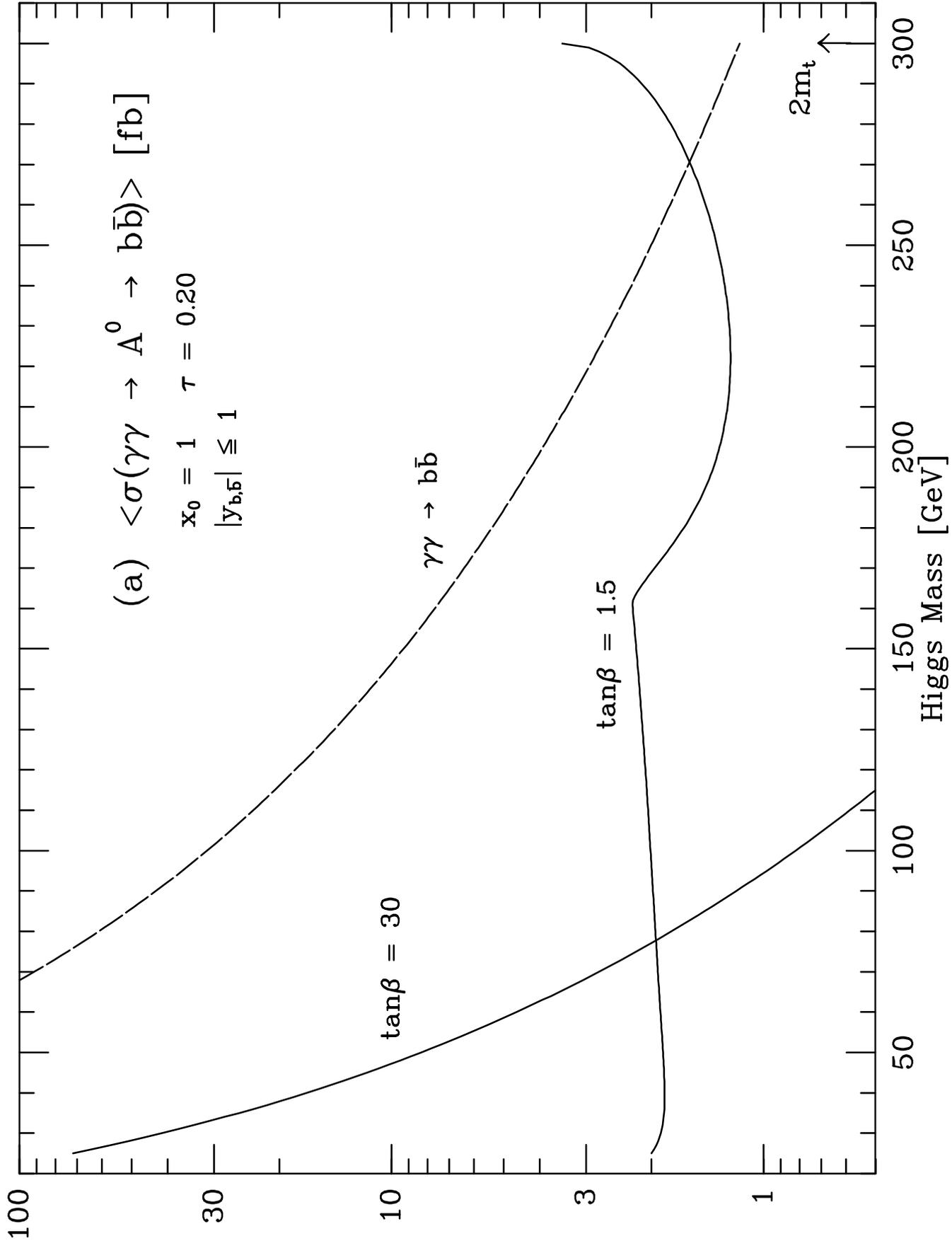
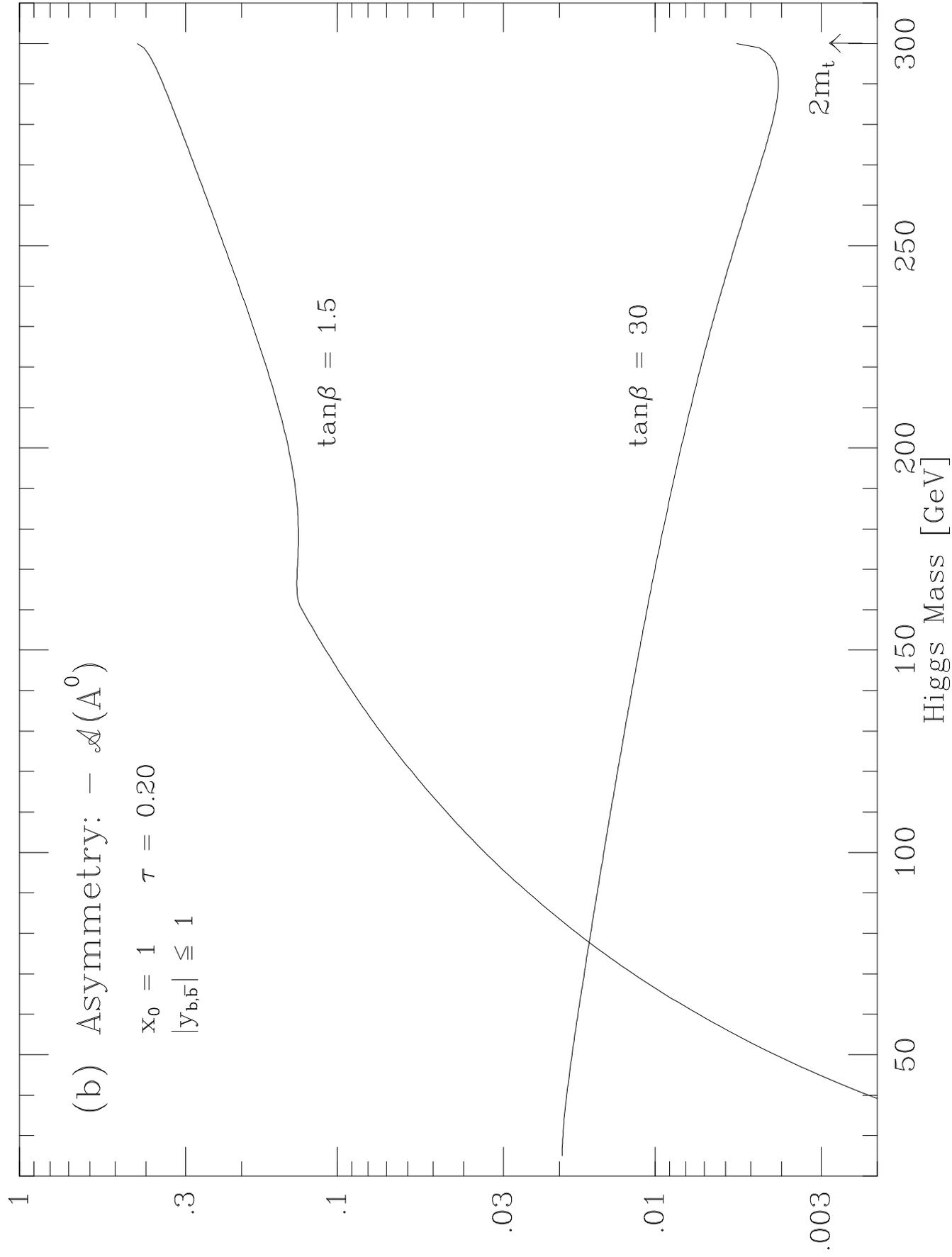
\end{document}